\documentclass[10pt]{IEEEtran}
\usepackage{amsmath}
\usepackage[T1]{fontenc}
\usepackage{algorithm}
\usepackage{alphalph}
\usepackage{algorithmic}
\usepackage{listings}

\usepackage{nomencl}
\makenomenclature

\usepackage{amsmath}
\usepackage{graphicx}
\usepackage{color}
\usepackage[cmintegrals]{newtxmath}
\usepackage[caption=false,font=footnotesize]{subfig}
\ifCLASSOPTIONcompsoc
\usepackage[caption=false,font=normalsize,labelfont=sf,textfont=sf]{subfig}
\else
\usepackage[caption=false,font=footnotesize]{subfig}
\fi
\usepackage[colorlinks,citecolor=red,urlcolor=blue,bookmarks=false,hypertexnames=true]{hyperref} 
\usepackage{caption}
\usepackage{lipsum}
\usepackage{nameref}
\usepackage{varioref}
\usepackage{hyperref}
\usepackage{setspace}

\newtheorem{lemma}{Lemma}
\newtheorem{proposition}{Proposition}

\begin{document}
\title{A Theoretically Novel Trade-off for Sparse Secret-key Generation}
\author{Makan~Zamanipour,~\IEEEmembership{Member,~IEEE}
\thanks{Manuscript received NOV, 2021; revised X XX, XXXX. Copyright (c) 2015 IEEE. Personal use of this material is permitted. However, permission to use this material for any other purposes must be obtained from the IEEE by sending a request to pubs-permissions@ieee.org. Makan Zamanipour is with Lahijan University, Shaghayegh Street, Po. Box 1616, Lahijan, 44131, Iran, makan.zamanipour.2015@ieee.org. }
}

\maketitle
\markboth{IEEE, VOL. XX, NO. XX, X 2021}%
{Shell \MakeLowercase{\textit{et al.}}: Bare Demo of IEEEtran.cls for Computer Society Journals}
\begin{abstract}
We in this paper theoretically go over a rate-distortion based sparse dictionary learning problem. We show that the Degrees-of-Freedom (DoF) interested to be calculated $-$ satnding for the minimal set that guarantees our rate-distortion trade-off $-$ are basically accessible through a \textit{Langevin} equation. We indeed explore that the relative time evolution of DoF, i.e., the transition jumps is the essential issue for a relaxation over the relative optimisation problem. We subsequently prove the aforementioned relaxation through the \textit{Graphon} principle w.r.t. a stochastic \textit{Chordal Schramm-Loewner} evolution etc {via a minimisation over a distortion between the relative realisation times of two given graphs $\mathscr{G}_1$ and $\mathscr{G}_2$ as $   \mathop{{\rm \mathbb{M}in}}\limits_{ \mathscr{G}_1, \mathscr{G}_2} {\rm \; }   \mathcal{D} \Big(  t\big(  \mathscr{G}_1 , \mathscr{G} \big)  ,  t\big(  \mathscr{G}_2 , \mathscr{G} \big)  \Big)$}. We also extend our scenario to the eavesdropping case. We finally prove the efficiency of our proposed scheme via simulations.
\end{abstract}


\maketitle

\IEEEdisplaynontitleabstractindextext
\IEEEpeerreviewmaketitle

\section{Introduction} 

\IEEEPARstart{M}any machine learning and modern signal processing applications $-$ such as bio-metric authentication/identification and recommending systems $-$ , follow sparse signal processing techniques \cite{1, 2, 3, 4, 5, 6}. The sparse synthesis model focuses on those data sets that can be approximated using a linear combination of only a small number of cells of a dictionary. The applications of sparse coding based dictionary learning go chiefly over either or both extraction and estimation of local features. Typically, this kind of discipline is controlled via a prior decomposition of the original signal into overlapping blocks $-$ call side-information. In relation to this strategy, there mainly exist two demerits: (\textit{i}) each point in the signal is estimated multiple times $-$ something that shows a redundancy; and (\textit{ii}) a shifted version of the features interested to be learnt are captured since the correlations/side-information among neighboring data sets are not fully taken into account $-$ something that results in the inevitable fact that some portions of the learnt dictionaries may be practically of a partially useless nature. 

A quick overview over the literature in terms of key-generation and authentication is strongly widely provided here in details. In \cite{bb1} and for cloud-assisted autonomous vehicles, Q. Jiang \textit{et al} technically proposed a bio-metric privacy preserving three-factor authentication and key agreement. In \cite{bb2} and through vehicular crowd-sourcing, F. Song \textit{et al} technically proposed a privacy-preserving task matching with threshold similarity search. In \cite{bb3} and for mobile Internet-of-Things (IoT), X. Zeng, \textit{et al} technically proposed a highly effective anonymous user authentication protocol. In \cite{bb4} and for fog-assisted IoT, J. Zhang \textit{et al} technically proposed a revocable and privacy-preserving decentralized data sharing scheme. In \cite{bb5} and for resource-constrained IoT devices, Q. Hu \textit{et al} technically proposed a half-duplex mode based secure key generation method. In \cite{bb6} and for mobile crowd-sensing under un-trusted platform, Z. Wang \textit{et al} theoretically explored privacy-driven truthful incentives. In \cite{bb7} and for IoT-enabled probabilistic smart contracts, N. S. Patel \textit{et al} theoretically explored blockchain-envisioned trusted random estimators. In \cite{bb8} and for IoT, R. Chaudhary \textit{et al} theoretically explored Lattice-based public key crypto-mechanisms. In \cite{bb9} and for vehicle to vehicle communication, V. Hassija \textit{et al} theoretically explored a scheme with the aid of directed acyclic graph and game theory. In \cite{bb10} and for vehicular social networks, J. Sun \textit{et al} theoretically explored a secure flexible and tampering-resistant data sharing mechanism. In \cite{bb11} and for mission-critical IoT applications, H. Wang \textit{et al} theoretically explored a secure short-packet communications. In \cite{bb12} and for edge-assisted IoT, P. Gope \textit{et al} theoretically proposed a highly effective privacy-preserving authenticated key agreement framework.

Overall, many works have been fulfilled in the context of secret key-generation, however, sparse coding and an optimised problem in this area has not been introduced so far. In addition, although the literature has tried its best to propose novel frameworks \cite{bb1, bb3, bb4, bb7, bb10, bb11, bb12} as well as the optimisation techniques \cite{bb2, bb5, bb6, bb8, bb9} and the relative relaxations, the population of the research and work done in this context still essentially lacks and requires to be further enhanced.

\begin{figure*}[t]
\centering
\subfloat{\includegraphics[trim={{11mm} {192 mm} {21mm} {23mm}},clip,scale=0.9]{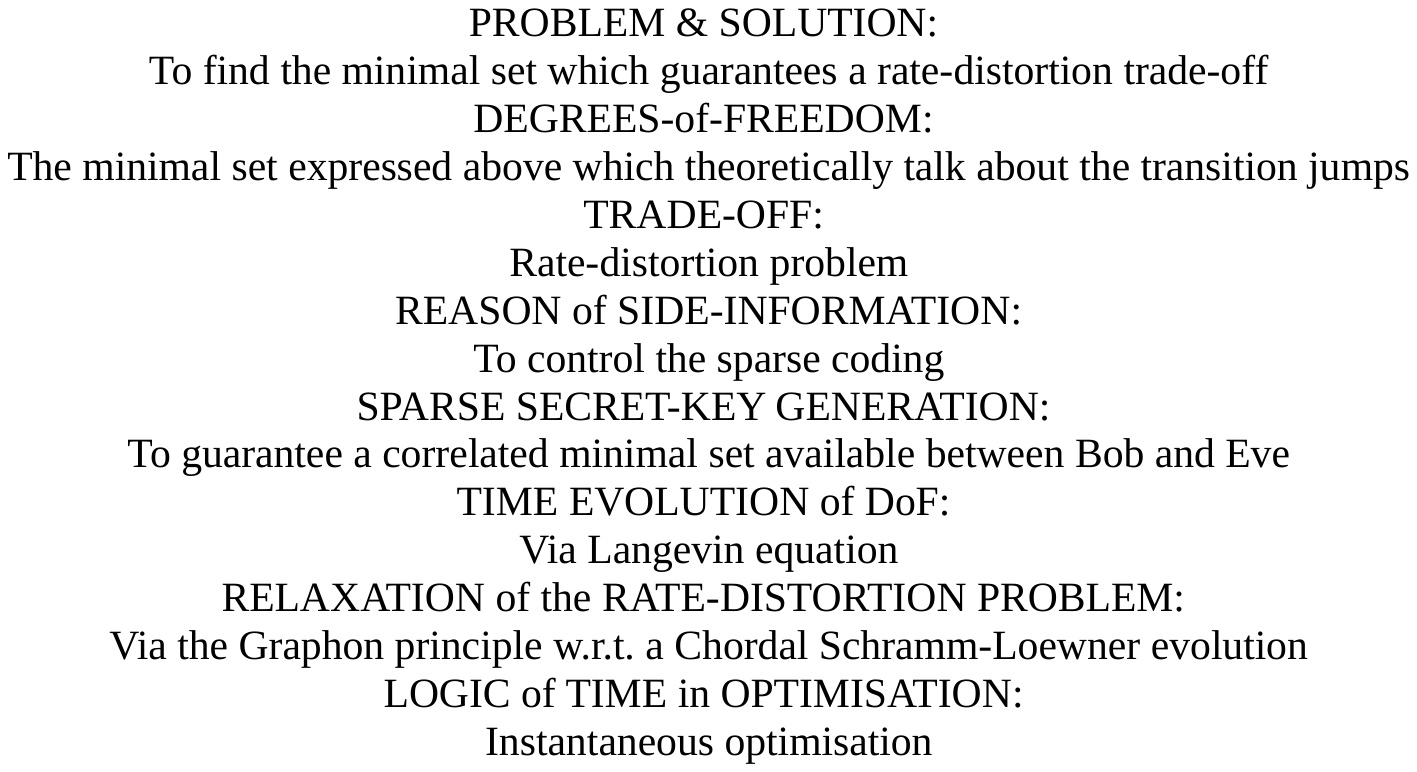}}
\caption{Essential details of our problem-and solution and its flow.} 
\label{F1}
\end{figure*} 

\subsection{Motivations and contributions} In this paper, we are interested in responding to the following question: \textit{How can we guarantee highly adequate relaxations over a rate-distortion based sparse dictionary learning? How is Alice able to efficiently utilise totally disparate resources to guarantee her private message to be concealed from Eve?} With regard to the non-complete version of the literature, the expressed question strongly motivate us to find an interesting solution, according to which our contributions are fundamentally described as follows. 
\begin{itemize}
\item \textcolor{red}{\textbf{(\textit{i})}} We initially write a rate-distortion based sparse coding optimisation problem. Subsequently, we try to add some extra constraints which are of a purely meaningful nature in this context. 

\item \textcolor{red}{\textbf{(\textit{ii})}} We show that we can get access to the time evolution of the Degrees-of-Freedom (DoF) in which we are interested. We do this through a \textit{Langevin} equation.

\item \textcolor{red}{\textbf{(\textit{iii})}} We then add an additional constraint in relation to the side-information.

\item \textcolor{red}{\textbf{(\textit{iv})}} Pen-ultimately, we make trials to theoretically relax the resultant optimisation problem w.r.t. the facts of \textit{Graphon} and a stochastic \textit{Chordal Schramm-Loewner} evolution etc. {$-$ something that can be chiefly elaborated according to some theoretical principles such as \textit{Riemannian gradients} in \textit{Riemannian geometry} and the \textit{Kroner's graph entropy}.}

\item \textcolor{red}{\textbf{(\textit{v})}} Finally, we also extend our scenario to the eavesdropping case. We in fact technically define some probabilistic constraints to be added to the main optimisation problem. 

\end{itemize}

\subsection{General notation} The notations widely used throughout the paper is given in Table \ref{table1}.

\begin{table*}[h]
\begin{center}
\captionof{table}{{List of notations.}}
 \label{table1} 
\begin{tabular}{p{2.091cm}|p{3.45cm} |p{3.26cm}|p{3.84cm}}
 \hline      
 \hline
Notation        &  Definition  & Notation          &  Definition   \\
\hline
 $\mathcal{P}_1$ &     Problem $1$ &    $\mathcal{P}_2$ &     Problem $2$        \\
$\mathcal{P}_3$ &     Problem $3$ &    $\mathcal{P}^{(\cdot)}_3$ &     New Version of Problem $3$        \\
    $\theta$   &                   Perturbation         &        $\mathbb{E}$ & Expected-value          \\
            $t$    &      Time Instance     &      $\mathscr{A}$      &  Side-information      \\
            $\sim$    &    Distributed       &      $\mathscr{W}_i, i \in \{1, 2\}$      &   Random-walk     \\
          $\xi$    &    Specific    Time Instance     &      $\mathscr{\lambda}_i, i \in \{1, 2, \cdots, 7\}$      &  Arbitary Threshold     \\
            $\mathcal{D} (\cdot)$    &     Distortion Function      &      $\mathscr{I} (\cdot)$      &  Mutual Information      \\
            $\mathscr{G}_i, i \in \{1, 2\}$    &    Graph       &      $\mathscr{B}^{(\cdot)}_i, i \in \{1, 2\}$      & Brownian Motion        \\
\hline
 \hline
\end{tabular}
\end{center}
\end{table*}

\subsection{Organisation}
The rest of the paper is organised as follows. The system set-up and our main results are given in Sections II and III. Subsequently, the evaluation of the framework and conclusions are given in Sections IV and V.

\section{System model and problem formulation}
In this section, we describe the system model, subsequently, we formulate the basis of our problem. 

\subsection{Flow of the problem-and-solution}

Initially speaking, take a quick look at Fig. \ref{F1}. It basically says:
\begin{itemize}
\item \textit{Problem-and-Solution} is to find the minimal set which guarantees a rate-distortion trade-off;
\item \textit{DoF} stands fundamentally theoretically for the minimal set expressed above which essentially talk about the transition jumps;
\item \textit{Trade-off} is the rate-distortion problem;
\item \textit{Reason of side-information} and the logic behind of using that is to control the sparse coding\footnote{Something that is traditional as mentioned in the paper in the next parts.};
\item \textit{Sparse secret-key generation} and the logic behind of using that is to guarantee a correlated minimal set available between Bob and Eve;
\item \textit{Time evolution of DoF} is explored via a \textit{Langevin} equation; 
\item \textit{Relaxation of the rate-distortion problem} is interpreted through the \textit{Graphon} principle and w.r.t. a stochastic \textit{Chordal Schramm-Loewner} evolution; and
\item \textit{Logic of time in optimisation} is to guarantee an instantaneous optimisation.
\end{itemize}

\subsection{System description}

A traditional sparse coding scheme practically includes the following steps \cite{1}:
\begin{itemize}
\item Sender: With the aid of a sparsifying transform which should be trained to the receiver $-$ call Bob $-$ via the server, the sender $-$ call Alice $-$ generates the sparse
code-words from her own data. She subsequently shares the privacy-protected sparse code-book with the the server as the service provider. Alice now decides\footnote{According to the Kerckchoffs’s principle in cryptography \cite{1}.} to permit a licensed availability, i.e., publicity over the sparsifying transform learnt by the client(s).
\item Server $-$ Step 1 as an indexing for the sender: The server marks some indexes on the received sparse-codes in a database.
\item Client: Via the shared sparsifying transform, Bob as the client generates a
sparse representation from his query data and subsequently sends it to the server.
\item Server $-$ Step 2 as a matching test for the client: The server searches to find the most similar sparse-code set, and subsequently replies to the client’s request.
\end{itemize}

In fact, Alice has a secret message while she decides to send it to Bob in the presence of the  passive $-$ potential $-$ or/and active $-$ actual $-$ Eve(s).

\subsection{Main problem}

A sparse dictionary learning problem can be mathematically formulated as \cite{1, 2, 3, 4, 5, 6}
\begin{equation}\label{eq:1}
\begin{split}
\;  \mathscr{P}_1:  \mathop{{\rm \mathbb{M}in}}\limits_{\mathscr{D},\Phi_0 \big(\mathscr{X}\big)  } {\rm \; } \mathop{{\rm \mathbb{E}}}\limits_{\theta \sim \Theta } \Bigg\{\mathcal{D} \Big(\mathscr{Y},\mathscr{D}\Phi_0 \big(\mathscr{X}; \theta \big) \Big)\Bigg\}, \\ s.t.\;\;\;\; |\mathscr{D}| \leq \lambda_0, \; \; \; \; \; \; \; \; \; \; \; \; \; \; \; \; \; \; \; \; \; \; \; \; \;  \\ \; |\Phi_0 \big(\mathscr{X}\big)| \leq \lambda_1, \; \; \; \; \; \; \; \; \; \; \; \; \;\; \; \; \; \; 
\end{split}
\end{equation}
with regard to the thresholds $\lambda_0$ and $\lambda_1$, where $\mathscr{Y}$ denotes the observed data, $\mathscr{D}$ represents the unknown dictionary, $\mathscr{X}$ stands for the original data set w.r.t. the data instance $x \in \mathbb{R}^{n}$, $\Phi_0 \big(\mathscr{X}\big)$ refers to the sparse representation coefficient set, $\Theta$ is a distribution over appropriate perturbations $\theta \sim \Theta$ \cite{6}, while we have the pre-trained models $\Phi: \mathbb{R}^{n} \rightarrow \mathbb{R}^{m}$ and $\Phi_0 \subset \Phi: \mathbb{R}^{n} \rightarrow \mathbb{R}^{m}$.

\textsc{\textbf{Definition 1\footnote{See e.g. \cite{r1, r2}.}: $\alpha-$secret key.}} \textit{A random variable $\mathcal{K}$ with the finite range $\mathscr{K}$ information-theoretically represents an $\alpha-$secret key for Alice and Bob, achievable through the public communication $\beta$, if there exist two functions $\mathscr{F}_a$ and $\mathscr{F}_b$ such that for any $\alpha >0$ the following three conditions hold: }
\begin{itemize}
\item \textit{(i) $\mathcal{P}r \Big(\mathcal{K}= \mathscr{F}_a \big(    \mathscr{X},\beta \big) = \mathscr{F}_b \big(    \mathscr{Y},\beta \big)  \Big) \ge 1- \alpha$; }
\item \textit{(ii) $\mathscr{I} \big ( \mathcal{K};\beta \big) \le \alpha$; and }
\item \textit{(iii) $\mathscr{H} \big ( \mathcal{K} \big) \ge log   |\mathscr{K}|  -\alpha$. }
\end{itemize}
\textit{These three conditions respectively ensure  that (i) Alice  and  Bob  generate  the  same  secret  key  with  high  probability as much as possible; the generated secret key is efficiently kept hide from any user except for Bob who gets access to the public communication $\beta$; and (iii) the secret key is sub-uniformly distributed.}

This definition is based fully upon a passive Eve which can be extended to an active scenario $-$ although we in the next parts, use the active one.

\section{Main results}
In this section, our main results are theoretically provided in details. 

\subsection{Time evolution of the sparsity}

\begin{proposition} \label{P2} \textit{The time evolution of the sparsity values, that is, as DoF relating to the transition jumps for the rate-distortion based sparse coding optimisation problem $\mathscr{P}_1$ expressed in Eqn. (\ref{eq:1}) can be mathematically attainable for the distortion term $\mathcal{D} \bigg( \Phi_1 \big(\mathscr{X}^{(t)}\big)  ,\Phi_0 \big(\mathscr{X}^{(t)}\big)\bigg)$ while $\Phi_1 \big(\mathscr{X}^{(t)}\big):=\Phi \Big( \Phi_0 \big(\mathscr{X}^{(t)}\big);\theta^{(t)},t \Big)$ for the instant $t$.}\end{proposition} 

\textbf{\textsc{Proof:}} See Appendix \ref{sec:A}.$\; \; \; \blacksquare$

\subsection{Relaxation of the rate-distorion problem}

\begin{proposition} \label{P2} \textit{While the degrees of freedom of interest so far is $\Phi_1 \big(\mathscr{X}^{(t)}\big)$, following the previous result, i.e., Proposition 1 and its relative proof given in Appendix \ref{sec:A}, the main problem over the distortion term $\mathcal{D} \bigg( \Phi_1 \big(\mathscr{X}^{(t)}\big)  ,\Phi_0 \big(\mathscr{X}^{(t)}\big)\bigg)$ is $NP-$hard. However, this problem can be theoretically relaxable through a new definition over the given side-information $\mathscr{A}$ as a new distortion constraint $\mathcal{D}\bigg(  \Phi_0 \big(\mathscr{X}^{(t)}\big) , \mathscr{A}^{(t)} \bigg)$.}\end{proposition} 

\textbf{\textsc{Proof:}} See Appendix \ref{sec:B}.$\; \; \; \blacksquare$

Algorithm \ref{fff} can solve $\mathscr{P}_2$ given in Eqn. (\ref{eq:6}) in Appendix \ref{sec:B}.

\begin{algorithm}
\caption{{A greedy algorithm to the Problem $ \mathscr{P}_2$.}}\label{fff}
\begin{algorithmic}
\STATE \textbf{\textsc{Initialisation.}} 

\textbf{while} $ \mathbb{TRUE}$ \textbf{do} 

$\;\;\;\;\;\;\;\;\;\;\;\;\;\;\;\;\;\;\;\;\;\;\;\;$(\textit{i}) Solve $\mathscr{P}_2$ given in Eqn. (\ref{eq:6}); and 

$\;\;\;\;\;\;\;\;\;\;\;\;\;\;\;\;\;\;\;\;\;\;\;\;$(\textit{ii}) Update. 

\textbf{endwhile} 

\textbf{\textsc{Output.}} 

\textbf{end}

\end{algorithmic}
\end{algorithm}

\subsection{Positive definity}

\begin{proposition} \label{P2} \textit{DoF, if they are non-zero, they are conditionally positive definite.}\end{proposition} 

\textbf{\textsc{Proof:}} See Appendix \ref{sec:C}.$\; \; \; \blacksquare$

\subsection{Extension to the eavesdropping scenraio}

Assume\footnote{See e.g. \cite{f5, f6}.} that we have two users, without loss of generality one of which is Bob as the legitimate user, and another one is an eavesdropper $-$ call Eve $-$ who can get access to the information in relation to the private message arose between Alice and Bob.

\begin{proposition} \label{P2} \textit{For the $t-$th instant, call the sparsity patterns $\mathcal{S}^{(t)}_{ab} \in \{ 0,1 \}$ and $\mathcal{S}^{(t)}_{ae} \in \{ 0,1 \}$ respectively for Bob and Eve, that is, in relation to the channel between Alice and Bob, and the one between Alice and Eve, respectively. For the $t-$th instant and $j-$th sub-channel, define\footnote{See e.g. \cite{f5, f6}.} $\mathcal{P}r \big( j \in \mathcal{S}^{(t)}_{ae}| j \in \mathcal{S}^{(t)}_{ab}\big)=\psi^{(t)}, \forall j \in \mathbb{N}$. The Problem $\mathscr{P}_2$ expressed in Eqn. (\ref{eq:6}) can be re-written as the Problem $\mathscr{P}_3$ expressed in Eqn. (\ref{eq:7}), or the Problem $\mathscr{P}^{\prime}_{3}$ expressed in Eqn. (\ref{eq:8}) or even the Problem $\mathscr{P}^{\prime\prime}_{3}$ expressed in Eqn. (\ref{eq:9}).}\end{proposition} 

\textbf{\textsc{Proof:}} See Appendix \ref{sec:D}.$\; \; \; \blacksquare$

\section{Numerical results}

We have done our simulations w.r.t. the Bernoulli-distributed data-sets using GNU Octave of version $4.2.2$
on Ubuntu $16.04$. 
 
\begin{figure}[t]
\centering
\subfloat{\includegraphics[trim={{20mm} {68 mm} {23mm} {73mm}},clip,scale=0.43]{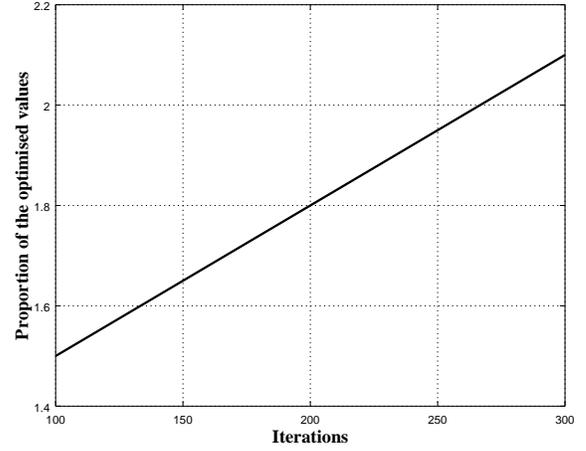}}
\caption{Instantaneous optimum value $ \Phi_0\big(\mathscr{X}^{(t)}\big)$ vs. iterations: How Algorithm \ref{fff} converges w.r.t. the thresholds $\lambda_2$ and $\lambda_3$ in the Problem $\mathscr{P}_2$ given in Eqn. (\ref{eq:6}).} 
\label{F3}
\end{figure}

Fig. \ref{F3} shows the optimum value earned from the Problem $\mathscr{P}_2$ given in Eqn. (\ref{eq:6}) while $\lambda_2$ and $\lambda_3$ are $\lll \infty$ divided by the one when $\lambda_2=\lambda_3=\infty$, i.e., the case in which we do not constrain our optimisation problem w.r.t. the relative constraints. As totally, obvious, the more constraints we intelligently and purposefully provide, the significantly more favourable response we can undoubtedly guarantee with regard to our proposed scheme. 

\begin{figure}[t]
\centering
\subfloat{\includegraphics[trim={{20mm} {68 mm} {23mm} {73mm}},clip,scale=0.43]{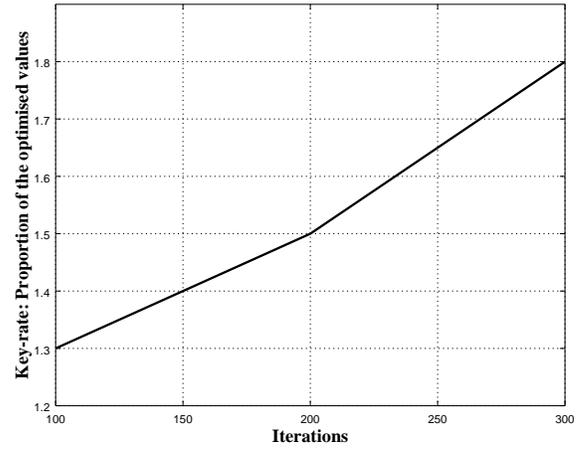}}
\caption{ Instantaneous key-rate $\mathscr{R}^{(t)}_{key} $ vs. iterations: How Algorithm \ref{fff} converges w.r.t. the thresholds $\lambda_2$ and $\lambda_3$ in the Problem $\mathscr{P}^{\prime\prime}_2$ given in Eqn. (\ref{eq:9}).} 
\label{F4}
\end{figure}

Fig. \ref{F4} shows the instantaneous key-rate $\mathscr{R}^{(t)}_{key}$ earned from the Problem $\mathscr{P}^{\prime\prime}_2$ given in Eqn. (\ref{eq:9}) while $\lambda_2$ and $\lambda_3$ are $\lll \infty$ divided by the one when $\lambda_2=\lambda_3=\infty$, i.e., the case in which we do not constrain our optimisation problem w.r.t. the relative constraints. The same as the previous trial, the more constraints we smartly provide, the considerably more adequate response we consequently assure in relation to our proposed scheme.

\begin{figure}[t]
\centering
\subfloat{\includegraphics[trim={{20mm} {68 mm} {23mm} {73mm}},clip,scale=0.43]{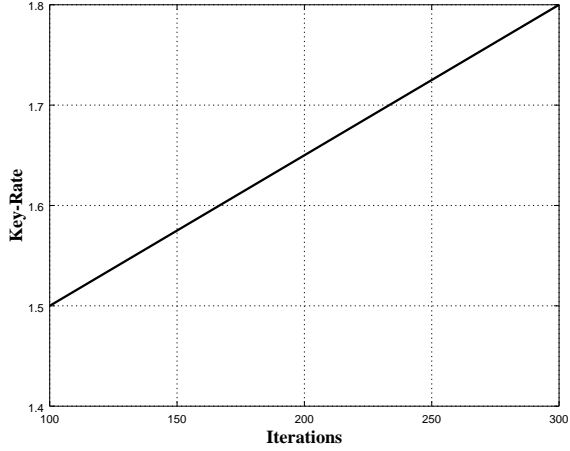}}
\caption{ Instantaneous key-rate $\mathscr{R}^{(t)}_{key} $ vs. iterations: Proported version of with to without the first constraint given in Eqn. (\ref{eq:4}), i.e., relating to the threshold $\lambda^{(t)}_1$.} 
\label{F7}
\end{figure}

Fig. \ref{F7} shows the instantaneous key-rate $\mathscr{R}^{(t)}_{key} $ against the iteration regime while it is proported version of the with case to the without one of the first constraint given in Eqn. (\ref{eq:4}), i.e., relating to the threshold $\lambda^{(t)}_1$.

\begin{figure}[t]
\centering
\subfloat{\includegraphics[trim={{20mm} {68 mm} {23mm} {73mm}},clip,scale=0.43]{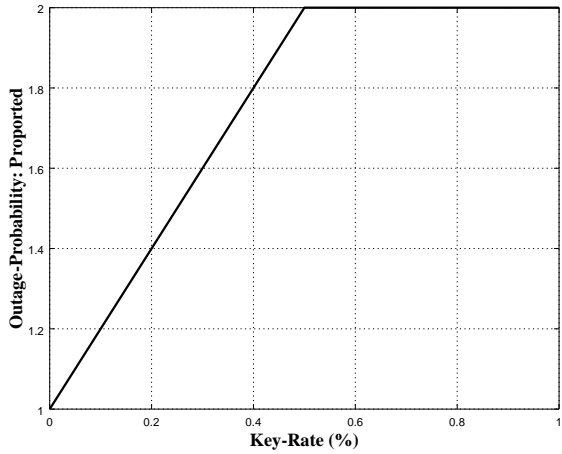}}
\caption{ Outage probablity $1-\lambda_9$ vs. the normalised version of the Key-Rate $\mathscr{R}^{(t)}_{key} $ given in Eqn. (\ref{eq:9}): Proported of $||\Delta \xi ||=15 \%$ to $||\Delta \xi ||=5 \%$.} 
\label{F8}
\end{figure}

Finally, the outage probablity $1-\lambda_9$ versus the normalised version of the Key-Rate $\mathscr{R}^{(t)}_{key} $ given in Eqn. (\ref{eq:9}) is shown in Fig. \ref{F8}  while it is proported of $||\Delta \xi ||=15 \%$ to $||\Delta \xi ||=5 \%$, where $||\cdot||$ stands for the Frobenius norm.

\section{conclusion}
We have theoretically shown a novel solution to an $NP-$hard rate-distortion based sparse dictionary learning in the context of a novel method of relaxation. We theoretically showed the accessibility of DoF in which er are interested via a \textit{Langevin} equation. In fact, the relative time evolution of DoF, i.e., the transition jumps for a relaxation over the relative optimisation problem was calculated in the context of a tractable fashion. The aforementioned relaxation was proven with the aid of the \textit{Graphon} principle w.r.t. a stochastic \textit{Chordal Schramm-Loewner} evolution etc. Our proposed method and solutions were extended to the eavesdropping case. The effectiveness of our proposed framework was ultimately shown by simulations.

\appendices
\section{Proof of Proposition 1}
\label{sec:A}
The problem $\mathscr{P}_1$ written in Eqn. (\ref{eq:1}) can be theoretically re-expressed as
\begin{equation}
\begin{split}
\;    \mathop{{\rm \mathbb{M}in}}\limits_{\Phi_0 \big(\mathscr{X}\big)} {\rm \; }   \mathop{{\rm \mathbb{E}}}\limits_{\theta \sim \Theta } \Bigg\{\mathcal{D} \bigg( \underbrace{\Phi \Big( \Phi_0 \big(\mathscr{X}\big);\theta \Big)}  _{\Phi_1 \big(\mathscr{X}\big)} ,\Phi_0 \big(\mathscr{X}\big)\bigg)\Bigg\}, \\ s.t.\;\;\;\; |\Phi_0 \big(\mathscr{X}\big)| \leq \lambda_1, \; \; \; \; \; \; \; \; \; \; \; \; \; \; \; \; \; \; \; \; \; \; \; \; \; \; \; 
\end{split}
\end{equation}
while $\Phi_1 \subset \Phi: \mathbb{R}^{n} \rightarrow \mathbb{R}^{m}$ for which $\Phi_1 \supset \Phi_0$ holds, or correspondingly in the context of an instantaneous\footnote{{See e.g. \cite{an1, an2} to understand what an instantaneous expected-value it means}.} optimisation problem
\begin{equation}
\begin{split}
\;    \mathop{{\rm \mathbb{M}in}}\limits_{\Phi_0 \big(\mathscr{X}^{(t)}\big)} {\rm \; }   \mathop{{\rm \mathbb{E}}}\limits_{\theta^{(t)} \sim \Theta } \Bigg\{\mathcal{D} \bigg( \underbrace{\Phi \Big( \Phi_0 \big(\mathscr{X}^{(t)}\big);\theta^{(t)},t \Big)}_{\Phi_1 \big(\mathscr{X}^{(t)}\big)}  ,\Phi_0 \big(\mathscr{X}^{(t)}\big)\bigg)\Bigg\}, \\ s.t.\;\;\;\; |\Phi_0 \big(\mathscr{X}^{(t)}\big)| \leq \lambda^{(t)}_1. \; \; \; \; \; \; \; \; \; \; \; \; \; \; \; \; \; \; \; \; \; \; \; \; \; \; \; \; \; \; \; \; \; \; \; \; \; 
\end{split}
\end{equation}
Now, as we are interested in theoretically exploring the sparsity, as DoF in relation to the transition jumps, one can apply the \textit{Langevin} equation according to which the time evolution of the aforementioned DoF can be accessible. Thus, we have
\begin{equation}\label{eq:4}
\begin{split}
\;    \mathop{{\rm \mathbb{M}in}}\limits_{\Phi_0 \big(\mathscr{X}^{(t)}\big)} {\rm \; }   \mathop{{\rm \mathbb{E}}}\limits_{\theta^{(t)} \sim \Theta } \Bigg\{\mathcal{D} \bigg( \underbrace{\Phi \Big( \Phi_0 \big(\mathscr{X}^{(t)}\big);\theta^{(t)},t \Big)}_{\Phi_1 \big(\mathscr{X}^{(t)}\big)}  ,\Phi_0 \big(\mathscr{X}^{(t)}\big)\bigg)\Bigg\}, \\ s.t.\;\;\;\; |\Phi_0 \big(\mathscr{X}^{(t)}\big)| \leq \lambda^{(t)}_1, \; \; \; \; \; \; \; \; \; \; \; \; \; \; \; \; \; \; \; \; \; \; \; \; \; \; \; \; \; \; \; \; \; \; \; \; \; \\
\frac{d \Big(\Phi_1 \big(\mathscr{X}^{(t)}\big)\Big)}{dt}=-\gamma^{(t)}_0 \Phi_1 \big(\mathscr{X}^{(t)}\big) + \mathscr{W}^{(t)}_1, \; \; \;  \; \; \; 
\end{split}
\end{equation}
while $\gamma^{(t)}_0$ is an arbitrary parameter, and the minus sign principally shows a backward orientation in the solution to the problem, as well as the theoretical fact that $\mathscr{W}^{(t)}_1$ is a random-walk.

The proof is now completed.$\; \; \; \blacksquare$

\section{Proof of Proposition 2}
\label{sec:B}
The proof is provided here in terms of the following multi-step solution. 

\textsc{Step 1.} It is strongly conventional \cite{10, 11} to define a constraint over a distortion between the side-information and the sparse data set. Therefore, we do this here as well according to which our optimisation problem is conveniently re-written as
\begin{equation}
\begin{split}
\;    \mathop{{\rm \mathbb{M}in}}\limits_{\Phi_0 \big(\mathscr{X}^{(t)}\big)} {\rm \; }   \mathop{{\rm \mathbb{E}}}\limits_{\theta^{(t)} \sim \Theta } \Bigg\{\mathcal{D} \bigg( \underbrace{\Phi \Big( \Phi_0 \big(\mathscr{X}^{(t)}\big);\theta^{(t)},t \Big)}_{\Phi_1 \big(\mathscr{X}^{(t)}\big)}  ,\Phi_0 \big(\mathscr{X}^{(t)}\big)\bigg)\Bigg\}, \\ s.t.\;\;\;\; |\Phi_0 \big(\mathscr{X}^{(t)}\big)| \leq \lambda^{(t)}_1, \; \; \; \; \; \; \; \; \; \; \; \; \; \; \; \; \; \; \; \; \; \; \; \; \; \; \; \; \; \; \; \; \; \; \; \; \; \\
\frac{d \Big(\Phi_1 \big(\mathscr{X}^{(t)}\big)\Big)}{dt}=-\gamma^{(t)}_0 \Phi_1 \big(\mathscr{X}^{(t)}\big) + \mathscr{W}^{(t)}_1, \; \; \;  \; \; \; \\
\mathcal{D}\bigg(  \Phi_0 \big(\mathscr{X}^{(t)}\big) , \mathscr{A}^{(t)} \bigg)\leq \lambda^{(t)}_2,\;\;\;\;\;\;\;\;\;\;\;\;\;\;\;\;\;\;\;\;\;\;\;\;
\end{split}
\end{equation}
w.r.t. the arbitary threshold $\lambda^{(t)}_2$.

\textsc{Step 2.} One can here say that DoF cited above can be in terms of the possibly available ways via which one can calculate a specific amount of information. Accordingly, the possibly available pieces of a specific amount of information calculable such as the relative side-information set can be DoF intersted here to be calculated. 


\textsc{Step 3.} Now, assume that the possibly available pieces of the specific amount of information calculable talked above are in the context of probabilistic energy\footnote{In order to see the relevance between energy and information see e.g. \cite{c1, c2, c3}.} quanta. Subsequently, we assume a given Ball $\mathscr{B}_0$ in which the number of energy quanta $\mathscr{e}$ should require
\begin{equation*}
\begin{split}
\; \mathbb{P}(\epsilon \leq \epsilon_0),
\end{split}
\end{equation*}
while $ \epsilon$ is in relation to $\mathscr{e}$, and we have a $2-d$ probability density function (pdf) $\mathscr{f}\big(\mathscr{e}, \mathscr{p}, t \big)$ while $\mathscr{e}$, $\mathscr{p}$ and $t$ stand for the energy, momentum and time, respectively. Subsequently, we assign  
\begin{equation*}
\begin{split}
\; d\mathscr{B}_0=d^2 \mathscr{e} d^2 \mathscr{p},
\end{split}
\end{equation*}
while $d^2$ is the second derivative in the polar axis. Thus, defining
\begin{equation*}
\begin{split}
\; d \Big(\Phi_1 \big(\mathscr{X}^{(t)}\big)\Big):= \mathscr{f}\big(\mathscr{e}, \mathscr{p}, t \big) d \mathscr{B}_0,
\end{split}
\end{equation*}
the relative energy quanta numbers is calculated as
\begin{equation*}
\begin{split}
\; \Phi_1 \big(\mathscr{X}^{(t)}\big):= \;\;\;\;\;\;\;\;\;\;\;\;\;\;\;\;\;\;\;\;\;\;\;\;\;\;\;\;\;\;\;\;\;\;\;\;\;\;\;\;\;\;\;\;\;\;\;\;\;\;\;\;\;\;\;\;\;\;\;\;\;\;\;\;\;\;\;\; \\ \oint_{\mathscr{B}_0}  \mathscr{f}\big(\mathscr{e}, \mathscr{p}, t \big) d \mathscr{B}_0=   \int_{\rho} \int_{\vartheta}  \mathscr{f}\big(\mathscr{e}, \mathscr{p}, t, \rho, \vartheta \big) \cdots \;\; \;\; \\  d \mathscr{e}_{\rho} (\rho,\vartheta,t) d \mathscr{e}_{\vartheta} (\rho,\vartheta,t) d \mathscr{p}_{\rho} (\rho,\vartheta,t) d \mathscr{p}_{\vartheta} (\rho,\vartheta,t).
\end{split}
\end{equation*}

\begin{figure*}
\begin{equation}\label{eq:6}
\begin{split}
\mathscr{P}_2: 
\begin{cases}
\;   \mathop{{\rm \mathbb{M}in}}\limits_{\Phi_0 \big(\mathscr{X}^{(t,\xi+\Delta \xi)}\big)} {\rm \; }   \mathop{{\rm \mathbb{E}}}\limits_{\theta^{(t,\xi+\Delta \xi)} \sim \Theta } \Bigg\{\mathcal{D} \bigg( \underbrace{\Phi \Big( \Phi_0 \big(\mathscr{X}^{(t,\xi+\Delta \xi)}\big);\theta^{(t,\xi+\Delta \xi)},t,\xi+\Delta \xi \Big)}_{\Phi_1 \big(\mathscr{X}^{(t,\xi+\Delta \xi)}\big)}  ,\Phi_0 \big(\mathscr{X}^{(t,\xi+\Delta \xi)}\big)\bigg)\Bigg\}, \\ 
s.t.\;\;\;\; 6.a: \; |\Phi_0 \big(\mathscr{X}^{(t,\xi+\Delta \xi)}\big)| \leq \lambda^{(t,\xi+\Delta \xi)}_1,  \\
6.b: \; \frac{d \Big(\Phi_1 \big(\mathscr{X}^{(t,\xi+\Delta \xi)}\big)\Big)}{dt}=-\gamma^{(t,\xi+\Delta \xi)}_0 \Phi_1 \big(\mathscr{X}^{(t,\xi+\Delta \xi)}\big) + \mathscr{W}^{(t,\xi+\Delta \xi)}_1, 	\exists \Phi^{(0)}_1 \big(\mathscr{X}^{(t_0)}\big),\\
6.c: \;\mathcal{D}\bigg(  \Phi_0 \big(\mathscr{X}^{(t,\xi+\Delta \xi)}\big) , \mathscr{A}^{(t,\xi+\Delta \xi)} \bigg)\leq \lambda^{(t,\xi+\Delta \xi)}_2,\\
6.d: \; t \in [t-\xi-\Delta \xi, t+\xi+\Delta \xi], 	\exists \xi>0,\exists \Delta \xi>0,\\
6.e: \; \mathcal{D}\bigg(  \Phi_0 \big(\mathscr{X}^{(t,\xi)}\big),  \Phi_0 \big(\mathscr{X}^{(t,\Delta \xi)}\big) \bigg)\leq \lambda^{(t)}_3.
\end{cases}
\end{split}
\end{equation}
\end{figure*}

\begin{figure*}
\begin{equation}\label{eq:7}
\begin{split}
\mathscr{P}_3: 
\begin{cases}
\;   \mathop{{\rm \mathbb{M}in}}\limits_{\Phi_0 \big(\mathscr{X}^{(t,\xi+\Delta \xi)}\big)} {\rm \; }   \mathop{{\rm \mathbb{E}}}\limits_{\theta^{(t,\xi+\Delta \xi)} \sim \Theta } \Bigg\{\mathcal{D} \bigg( \Phi_1 \big(\mathscr{X}^{(t,\xi+\Delta \xi)}\big) ,\Phi_0 \big(\mathscr{X}^{(t,\xi+\Delta \xi)}\big)\bigg)\Bigg\}, \\ s.t.\;\;\;\; 6.a-6.e,\\
 7.f: \; \mathcal{P}r \Big (\psi^{(t,\xi + \Delta \xi)} \big(  \mathscr{X}^{(t,\xi + \Delta \xi)}_0   \big) \ge \lambda^{(t,\xi + \Delta \xi)}_4 \Big) \le \lambda^{(t,\xi + \Delta \xi)}_5. 
\end{cases}
\end{split}
\end{equation}
\end{figure*}

\begin{figure*}
\begin{equation}\label{eq:8}
\begin{split}
\mathscr{P}^{\prime}_3: 
\begin{cases}
\;   \mathop{{\rm \mathbb{M}in}}\limits_{\Phi_0 \big(\mathscr{X}^{(t,\xi+\Delta \xi)}\big)} {\rm \; }   \mathop{{\rm \mathbb{E}}}\limits_{\theta^{(t,\xi+\Delta \xi)} \sim \Theta } \Bigg\{\mathcal{D} \bigg( \Phi_1 \big(\mathscr{X}^{(t,\xi+\Delta \xi)}\big) ,\Phi_0 \big(\mathscr{X}^{(t,\xi+\Delta \xi)}\big)\bigg)\Bigg\}, \\ s.t.\;\;\;\; 6.a-6.e, 7.f, \\
 8.h: \; \mathcal{P}r \Big (\omega^{(t,\xi + \Delta \xi)} \big(  \mathscr{X}^{(t,\xi + \Delta \xi)}_0   \big) \ge \lambda^{(t,\xi + \Delta \xi)}_6 \Big) \ge \lambda^{(t,\xi + \Delta \xi)}_7.
\end{cases}
\end{split}
\end{equation}
\end{figure*}

\begin{figure*}
\begin{equation}\label{eq:9}
\begin{split}
\mathscr{P}^{\prime\prime}_3: 
\begin{cases}
\;   \mathop{{\rm \mathbb{M}in}}\limits_{\Phi_0 \big(\mathscr{X}^{(t,\xi+\Delta \xi)}\big)} {\rm \; }   \mathop{{\rm \mathbb{E}}}\limits_{\theta^{(t,\xi+\Delta \xi)} \sim \Theta } \Bigg\{\mathcal{D} \bigg( \Phi_1 \big(\mathscr{X}^{(t,\xi+\Delta \xi)}\big)  ,\Phi_0 \big(\mathscr{X}^{(t,\xi+\Delta \xi)}\big)\bigg)\Bigg\}, \\ s.t.\;\;\;\; 6.a-6.e, \\
 9.f: \; \mathcal{P}r \Big (\mathscr{R}^{(t,\xi + \Delta \xi)}_{key} \big(  \mathscr{X}^{(t,\xi + \Delta \xi)}_0   \big) \ge \lambda^{(t,\xi + \Delta \xi)}_8 \Big) \ge \lambda^{(t,\xi + \Delta \xi)}_9.
\end{cases}
\end{split}
\end{equation}
\end{figure*}

\textsc{Step 4.} Now, theoretically imagine a game and the relative users as follows. The Reward is each of the possibly available pieces of the specific amount of information calculable in the previous step, for each relative DoF as the users. 

\textsc{\textbf{Remark 1.}} \textit{In our game, the population physically has a positive growth rate. This is because of the fact that the principle of Diffusion-and-Reaction hold here, that is, a diffusive flow exists from the side of a higher concentration to a less one, and reciprocally, a reactive movement exists from the less concentration towards the higher one.}

\textsc{Step 5.} Now, let us mathematically define an initial condition for the Langevin based constraint. We have $\Phi^{(0)}_1 \big(\mathscr{X}^{(t_0)}\big)$. Now, call the \textit{Picard-Lindelof} Theorem\footnote{See e.g. \cite{12} to see what it is.} which can be of a purely useful nature to say that: the Langevin equation 
\begin{equation*}
\begin{split}
\frac{d \Big(\Phi_1 \big(\mathscr{X}^{(t)}\big)\Big)}{dt}=-\gamma^{(t)}_0 \Phi_1 \big(\mathscr{X}^{(t)}\big) + \mathscr{W}^{(t)}_1,
\end{split}
\end{equation*}
w.r.t. the initial condition $\Phi^{(0)}_1 \big(\mathscr{X}^{(t_0)}\big)$ can have a unique solution $\Phi^{(*)}_1 \big(\mathscr{X}^{(t_*)}\big)$ w.r.t. a surely existing $ \xi >0$ over the zone $[t_0 - \xi , t_0 + \xi]$; while $\Phi_1 \big(\mathscr{X}^{(t}\big)$ is uniformly Lipschitz continuous in $\Phi_0 \big(\mathscr{X}^{(t}\big)$ and continuous in $t$. Thus, we indeed branch and localise the optimisation problem into its localities, without loss of generality and optimality. This is because of the $NP-$hard structure of our main optimisation problem since that is not convex, so, the strong duality does not theoretically hold \cite{e1}.

\textsc{Step 6.} Re-call \textit{Remark 1} for the fact that the population size of the relative users has a growth over the time zone, according to which we can theoretically consider a \textit{Graphon}\footnote{See e.g. \cite{6}.} in between. Indeed, we can prove that there exists a duality between the main optimisation problem and its localised version created in Step 5. Now, w.r.t. the fact that there would exist a stochastic \textit{Chordal}\footnote{Which maps a boundary point to another one, which is in contrast to the \textit{Radial Schramm-Loewner} evolution which maps a boundary point to an interior one: See e.g. \cite{13, 14} to see what they are.} \textit{Schramm-Loewner} evolution\footnote{See e.g. \cite{13, 14} to see what it is.} which shows a Brownian motion over the surface of the relative evolving Riemannian-manifold relating to the graph as
\begin{equation*}
\begin{split}
\; \lim_{\Delta \xi \rightarrow 0} \frac{\Phi_0 \big(\mathscr{X}^{(t,\xi+\Delta \xi)}\big)-\Phi_0 \big(\mathscr{X}^{(t,\xi)}\big)}{\Delta \xi}=\;\;\;\;\;\;\;\; \; \; \; \; \; \; \; \; \; \; \; \; \; \; \; \; \; \; \; \; \; \; \\ \mathcal{F}\Big(  \frac{1}{ \Phi_0 \big(\mathscr{X}^{(t,\xi)}\big), \mathscr{W}^{(t,\xi)}_2}  \Big),
\end{split}
\end{equation*}
w.r.t. the arbitary function $\mathcal{F}(\cdot)$ and the random-walk $ \mathscr{W}^{(t,\xi)}_2$. On other hand, since a graphon $-$ as a density measure $-$ is the number of times $\mathscr{G}_1$ occurs as a sub-graph of the graph $\mathscr{G}$, and w.r.t. the fact that two graphs, do they have the final approximation in terms of a same graphon, they actualise the same phenomenon \cite{s}, and regarding the fact that the positive growth of the population size literally defined above highly requires an adjacent matrix for the evolving Riemannian-manifold and the graph in between, a proof for a the duality we are looking for can be proven. Now, the duality basically holds iff and only iff $\mathscr{G}_1 \in \mathscr{G}$ and $\mathscr{G}_2 \in \mathscr{G} $ conclude a common phenomenon, while $\mathscr{G}_2$ fundamentally is a stochastic Schramm-Loewner evolutioned version of $\mathscr{G}_1$ in the nearest physical vicinity of the Reimannian sub-manifold, relating to the surface of the relative evolving Riemannian-manifold, that is
\begin{equation*}
\begin{split}
\; \mathcal{D} \Big(  t\big(  \mathscr{G}_1 , \mathscr{G} \big)  ,  t\big(  \mathscr{G}_2 , \mathscr{G} \big)  \Big) \rightarrow \nu_1^{+},
\end{split}
\end{equation*}
while $\nu_1^{+}$ is a sufficiently small non-zero value and $t\big(  \mathscr{G}_1 , \mathscr{G} \big)$ and $t\big(  \mathscr{G}_2 , \mathscr{G} \big)$ respectively show the final realisation of the relative graphs $\mathscr{G}_1$ and $\mathscr{G}_2$. This is equivalent to\footnote{This kind of optimisation problem can be solved by the \textit{alternating direction method of multipliers}: See e.g. \cite{c1, c2, c3}.}
\begin{equation*}
\begin{split}
\;   \mathop{{\rm \mathbb{M}in}}\limits_{ \mathscr{G}_1, \mathscr{G}_2} {\rm \; }   \mathcal{D} \Big(  t\big(  \mathscr{G}_1 , \mathscr{G} \big)  ,  t\big(  \mathscr{G}_2 , \mathscr{G} \big)  \Big).
\end{split}
\end{equation*}
Towards such end, we re-express our optimisation problem as Eqn. (\ref{eq:6}) on the top of the next page in which the term $\xi+\Delta \xi$ indicates the stochastic \textit{Schramm-Loewner} evolution expressed above and $\lambda^{(t)}_3$ is an arbitary threshold to constraint the amount of the \textit{Schramm-Loewner} evolution. {Now, let us continue by the following definition.}

{\textsc{\textbf{Definition 2\footnote{See e.g. \cite{koon1}.}: The $\mathscr{d}-$order binomial extension.}} \textit{Consider the mapping $\mathcal{V}: \mathcal{A} \longrightarrow \mathcal{B}$ $-$ such that $\mathcal{B}=\mathcal{V} \Big( \mathcal{A} \Big) $ holds $-$  and a $\mathscr{d} $ which is a discrete random variable that takes values on the non-zero non-negative integers $\mathbb{Z}^{+} \setminus \{ 0\}$. The mapping $\mathcal{V}^{\oplus \mathscr{d}}: \mathcal{A} \longrightarrow \mathcal{B}^{\mathscr{d}}$ $-$ such that $\mathcal{B}^{\mathscr{d}}=\mathcal{V}^{\oplus \mathscr{d}} \Big( \mathcal{A} \Big) $ holds $-$ is theoretically called the $\mathscr{d}-$order binomial extension of $\mathcal{V}$ if }
\begin{equation*}
\begin{split}
\;   \mathcal{V}^{\oplus \mathscr{d}} [\mathscr{b}^{\mathscr{d}} |\mathscr{a}]= \prod_{i=0}^{\mathscr{d}-1} \mathcal{V} [\mathscr{b}_i |\mathscr{a}] , \forall \mathscr{a} \in \mathcal{A}, \mathscr{b}^{\mathscr{d}} \in \mathcal{B}^{\mathscr{d}}.
\end{split}
\end{equation*}

{In the subsequence of \textit{Definition 2}, one can re-consider the optimisation}
\begin{equation*}
\begin{split}
{\;   \mathop{{\rm \mathbb{M}in}}\limits_{ \mathscr{G}_1, \mathscr{G}_2} {\rm \; }   \mathcal{D} \Big(  t\big(  \mathscr{G}_1 , \mathscr{G} \big)  ,  t\big(  \mathscr{G}_2 , \mathscr{G} \big)  \Big),}
\end{split}
\end{equation*}
as
\begin{equation*}
\begin{split}
\; {  \mathop{{\rm \mathbb{M}in}}\limits_{ \mathscr{G}_1, \mathcal{P}\big( \mathscr{d}\big)} {\rm \; }   \mathcal{D} \bigg(  t\big(  \mathscr{G}_1 , \mathscr{G} \big)  ,  t\Big( \underbrace{ \mathscr{G}_2= \mathcal{V}^{\oplus \mathscr{d}} \big(\mathscr{G}_1 \big)}_{ \approx \mathscr{G}^{\oplus \mathscr{d}}_1} , \mathscr{G} \Big)  \bigg),}
\end{split}
\end{equation*}
{while $\mathcal{P}\big( \mathscr{d}\big)$ basically is the probability mass function (pmf) of $\mathscr{d}$. It is also importantly interesting to be noticed that the term $(\cdot) ^{\oplus \mathscr{d}}$ comes principally from the convolutional nature of the Graph $ \mathscr{G}_2$ associated fundamentally with $ \mathscr{G}_1$ since the terms $t\big(  \mathscr{G}_1 , \mathscr{G} \big)  $ and $  t\big(  \mathscr{G}_2 , \mathscr{G} \big)$ may be technically of a Poisson-randomly-distributed nature. Meanwhile, we may interchangeably use $\approx \mathscr{G}^{\oplus \mathscr{d}}_1$ with a slight abuse of notation here, without loss of generality. }

{Now, consider the graph $\mathscr{G}$ and the subgraphs $\mathscr{G}_1$ and $\mathscr{G}_2$ with respectively random vertex sets $\mathcal{X}$, $\mathcal{X}_1$ and $\mathcal{X}_2$. Let $\mathcal{P}$, $\mathcal{P}_1$ and $\mathcal{P}_2$ be the probability density distributions respectively on $\mathcal{X}$, $\mathcal{X}_1$ and $\mathcal{X}_2$
whose marginal on $\mathcal{X}$ are equal to $\mathcal{P}$. Now the later optimisation can be re-written as}
\begin{equation*}
\begin{split}
\; {  \mathop{{\rm \mathbb{M}in}}\limits_{ \mathscr{G}_1, \mathcal{P}\big( \mathscr{d}\big)} {\rm \; }   \mathcal{D} \bigg(  t\big(  \mathscr{G}_1 , \mathscr{G} \big)  ,  t\Big( \underbrace{ \mathscr{G}_2= \mathcal{V}^{\oplus \mathscr{d}} \big(\mathscr{G}_1 \big)}_{ \approx \mathscr{G}^{\oplus \mathscr{d}}_1} , \mathscr{G} \Big)  \bigg),}\\
{    s.t. \;          \mathop{{\rm \mathbb{M}in}}\limits_{ \mathcal{P}_1} {\rm \; }    \mathscr{I} \big(    \mathcal{X},\mathcal{X}_1    \big)   =\mathscr{f}_1 \Big(   \mathcal{P}\big( \mathscr{d}\big) \Big)     ,         \;\;\;\;\;\;\;\;\;\;     }\\
{             \mathop{{\rm \mathbb{M}in}}\limits_{ \mathcal{P}_2} {\rm \; }    \mathscr{I} \big(    \mathcal{X},\mathcal{X}_2    \big)      =\mathscr{f}_2 \Big(   \mathcal{P}\big( \mathscr{d}\big) \Big)    ,           \;\;\;\;\;\;\;\;\;\;      }\\
{   \mathcal{D} \bigg(     \mathscr{f}_1 \Big(   \mathcal{P}\big( \mathscr{d}\big) \Big),\mathscr{f}_2 \Big(   \mathcal{P}\big( \mathscr{d}\big) \Big) \bigg) \le \nu_2^{+}   ,          \;\;\;\;    }\\
{\mathcal{X}_2= \mathcal{V}^{\oplus \mathscr{d}} \big(\mathcal{X}_1 \big) \approx \mathcal{X}^{\oplus \mathscr{d}}_1 ,      \;\;\;\;\;\;\;        \;\;\;\;\;\;\;\;\;\;   \;   }\\
{\mathscr{f}_2= \mathcal{V}^{\oplus \mathscr{d}} \big(\mathscr{f}_1 \big)  \approx \mathscr{f}^{\oplus \mathscr{d}}_1 ,     \;\;\;\;\;\;\;\;      \;\;\;\;\;\;\;\;\;\;    \;  }\\
\end{split}
\end{equation*}
where $\mathcal{P}:=\mathcal{P} \big(  \mathscr{d}\big)$ holds, w.r.t. the non-zero non-negative threshold $\nu_2^{+}$ as well as arbitary functions $ \mathscr{f}_1$ and $ \mathscr{f}_2$, while the first two additional constraints are thoroughly justified according to the \textit{Korner’s graph entropy}\footnote{\cite{hosh1, hosh2}.} $-$ something that can be stongly followed up via \textit{Riemannian gradients} in \textit{Riemannian geometry} as described below. Indeed, the first two-constraints have come from the fact that 
\begin{equation*}
\begin{split}
\; \mathop{{\rm \mathbb{M}in}}\limits_{ \mathscr{G}_1, \mathcal{P}\big( \mathscr{d}\big)} {\rm \; }   \mathcal{D} \bigg(  t\big(  \mathscr{G}_1 , \mathscr{G} \big)  ,  t\big(  \mathscr{G}^{\oplus \mathscr{d}}_1 , \mathscr{G} \big)  \bigg),
\end{split}
\end{equation*}
is equivalent to
\begin{equation*}
\begin{split}
\; { \mathop{{\rm \mathbb{M}in}}\limits_{ \cdot} {\rm \; }    \mathscr{I} \big(    \mathcal{X},\mathcal{X}_1| \mathcal{X}_2   \big)   - \mathscr{I} \big(    \mathcal{X},\mathcal{X}_2| \mathcal{X}_1   \big)\;\;\;\;\;\;\;\;\;\;\;\;\;\;\;\;\;\;\;\;\;\;\;\;\;\;\;\;\;\;\;\;\;\;\;\;}\\
{=\mathop{{\rm \mathbb{M}in}}\limits_{ \cdot} {\rm \; }    \mathscr{I} \big(    \mathcal{X}| \mathcal{X}_2   \big)   - \mathscr{I} \big(    \mathcal{X}|\mathcal{X}_1, \mathcal{X}_2  \big)-    \mathscr{I} \big(    \mathcal{X}|\mathcal{X}_1   \big)  + \mathscr{I} \big(    \mathcal{X}|\mathcal{X}_2, \mathcal{X}_1   \big)}\\
{ \equiv  \mathop{{\rm \mathbb{M}in}}\limits_{ \cdot} {\rm \; }    \mathscr{I} \big(    \mathcal{X}| \mathcal{X}_2   \big)   - \mathscr{I} \big(    \mathcal{X}|\mathcal{X}_1   \big)  \;\;\;\;\;\;\;\;\;\;\;\;\;\;\;\;\;\;\;\;\;\;\;\;\;\;\;\;\;\;\;\;\;\;\;\;\;\;\;\;\;\;\;\;\;\;}\\
{=\mathop{{\rm \mathbb{M}in}}\limits_{ \cdot} {\rm \; }    \mathscr{I} \big(    \mathcal{X}, \mathcal{X}_2   \big)   - \mathscr{I} \big(     \mathcal{X}_2  \big)-    \mathscr{I} \big(    \mathcal{X},\mathcal{X}_1   \big)  + \mathscr{I} \big(  \mathcal{X}_1   \big)\;\;\;\;\;\;\;\;\;\;\;\;\;\;\;\;\;}\\
{ \equiv \mathop{{\rm \mathbb{M}in}}\limits_{ \cdot} {\rm \; }    \mathscr{I} \big(    \mathcal{X}, \mathcal{X}_2   \big)    -    \mathscr{I} \big(    \mathcal{X},\mathcal{X}_1   \big) .\;\;\;\;\;\;\;\;\;\;\;\;\;\;\;\;\;\;\;\;\;\;\;\;\;\;\;\;\;\;\;\;\;\;\;\;\;\;\;\;\;\;\;\;}
\end{split}
\end{equation*}

The strongly theoretical logic behind of our interpretation and consideration over the evolution expressed above can be physically justified as follows as well, according to the \textit{Riemannian geometry} and \textit{Riemannian optimisation} principles\footnote{ See e.g. \cite{aa1, aa2, aa3}.}. Let us initially define the \textit{Riemannian gradients}\footnote{ See e.g. \cite{aa1, aa2, aa3, aa4, aa5}.} in relation to the sub-manifolds as the orthogonal \textit{projection} of the distortion function
\begin{equation}
\begin{split}
\;    \mathcal{D} \Big(  t\big(  \mathscr{G}_1 , \mathscr{G} \big)  ,  t\big(  \mathscr{G}_2 , \mathscr{G} \big)  \Big),
\end{split}
\end{equation}
onto the associated \textit{tangent space} as
\begin{equation*}
\begin{split}
\;   \mathcal{G}rad_{\mathscr{Rie}}:= \mathbb{P}roj _{ \mathscr{G}_2} \Big(  \nabla  \mathscr{G}_1  \Big),
\end{split}
\end{equation*}
which can be widely re-written in terms of the Riemannian logarithmic map operator 
\begin{equation*}
\begin{split}
\;   \mathcal{G}rad_{\mathscr{Rie}} = \Sigma^{\frac{1}{2}}_2 log \Big (    \Sigma^{-\frac{1}{2}}_2       \Sigma_2        \Sigma^{-\frac{1}{2}}_2 \Big) \Sigma^{\frac{1}{2}}_2,
\end{split}
\end{equation*}
while $\Sigma_1$ and $\Sigma_2$ theoretically are the eigen-value matrices associated with the graphs $\mathscr{G}_1$ and $\mathscr{G}_2$, respectively. Now, defining translation functions $\mathcal{T}_1$ and $\mathcal{T}_2$ respectively for the real-parts of $\Sigma_1$ and $\Sigma_2$, and rotation functions $\mathcal{R}^{(\mathscr{o})}_1$ and $\mathcal{R}^{(\mathscr{o})}_2$ for the relative imaginary-parts, while the following conditions satisify
\begin{equation*}
\begin{split}
\;  \mathcal{T}_i=\bar{\mathcal{T}_i}+\mathscr{B}^{(\mathcal{T})}_i, i \in \{ 1, 2\}, \;\;\;\;\;\;\;\;  \\
\mathcal{R}^{(\mathscr{o})}_i=\bar{\mathcal{R}}^{(\mathscr{o})}_i+\mathscr{B}^{(\mathcal{R})}_i, i \in \{ 1, 2\},
\end{split}
\end{equation*}
while $\bar{(\cdot)}$ principally indicates the average value and $\mathscr{B}^{(\mathcal{T})}_i, i \in \{ 1, 2\}$ and $\mathscr{B}^{(\mathcal{R})}_i, i \in \{ 1, 2\}$ are Brownian motions. Now, our relaxation we are focusing on is equivalent ot the following optimisation
\begin{equation*}
\begin{split}
\; \mathop{{\rm \mathbb{M}in}}\limits_{\mathcal{T}_1,\mathcal{T}_2, \mathcal{R}^{(\mathscr{o})}_1, \mathcal{R}^{(\mathscr{o})}_2} {\rm \; }   \mathcal{D} \Bigg( \mathcal{D} \bigg (  \mathcal{T}_1, \mathcal{T}_2 \bigg) , \mathcal{D} \bigg (  f \big( \mathcal{T}_1 \big), f \big(\mathcal{T}_2\big) \bigg) \Bigg)   +   \;\;\;\;\;\;\;\;\;\;\;\;\;\; \\  \mathcal{D} \Bigg( \mathcal{D} \bigg ( \mathcal{R}^{(\mathscr{o})}_1,\mathcal{R}^{(\mathscr{o})}_2 \bigg) , \mathcal{D} \bigg (  f \big( \mathcal{R}^{(\mathscr{o})}_1 \big), f \big(\mathcal{R}^{(\mathscr{o})}_2\big) \bigg) \Bigg)   .
\end{split}
\end{equation*}
\begin{lemma} \label{P2} \textit{For the equation derived above, the necessary and sufficient condition to be relaxable fundamentally is the fact that there should exist $\eta^{\mathcal{T}}$ and $ \eta^{\mathcal{R}^{(\mathscr{o})}}$ such that the following satisfies}\end{lemma}
\begin{equation*}
\begin{split}
\;    \mathcal{D} \bigg (  f \big( \mathcal{T}_1 \big), f \big(\mathcal{T}_2\big) \bigg)= \eta^{\mathcal{T}} \mathcal{D} \bigg (  \mathcal{T}_1, \mathcal{T}_2 \bigg)    ,  \;\;\;\;\;\;\;\;\;\;\;\;\;\;\;\;\;\;\; \\  \mathcal{D} \bigg (  f \big( \mathcal{R}^{(\mathscr{o})}_1 \big), f \big(\mathcal{R}^{(\mathscr{o})}_2\big) \bigg) =  \eta^{\mathcal{R}^{(\mathscr{o})}}  \mathcal{D} \bigg ( \mathcal{R}^{(\mathscr{o})}_1,\mathcal{R}^{(\mathscr{o})}_2 \bigg)    .
\end{split}
\end{equation*}

\textsc{\textbf{Proof:}} The proof is omitted for brevity.$\; \; \; \blacksquare$

\textsc{Step 7.} Finally, the following lemma can help us in concluding our interpretation.
 
\begin{lemma} \label{P2} \textit{The constraint 
\begin{equation*}
\begin{split}
\;   \mathcal{D}\bigg(  \Phi_0 \big(\mathscr{X}^{(t,\xi)}\big),  \Phi_0 \big(\mathscr{X}^{(t,\xi+\Delta \xi)}\big) \bigg)\leq \lambda^{(t)}_3,
\end{split}
\end{equation*}
is interpreted according to the Bochner's theorem.}\end{lemma} 

\textsc{\textbf{Proof:}} The Bochner's theorem\footnote{see e.g. \cite{p2} to see what it is.} says that for a locally compact abelian group\footnote{A group while each pair of group members are interchangible.} $\mathscr{g}$, with dual group $\hat{\mathscr{g}}$,  there exists a unique probability measure $\kappa$ such that 
\begin{equation*}
\begin{split}
\; \bar{\mathscr{f}}  \big(  \mathscr{g} \big) =\int_{\hat{\mathscr{g}}} \mathscr{x}  (  \mathscr{g} )d \kappa (\mathscr{x}),
\end{split}
\end{equation*}
holds. This completes the proof of \textit{Lemma 1} here.

The proof is now completed.$\; \; \; \blacksquare$

\section{Proof of Proposition 3}
\label{sec:C}
The \textit{Herglotz-Riesz} representation theorem for \textit{harmonic} functions\footnote{see e.g. \cite{p1} to see what it is.} can be useful here for us saying that DoF we are interested in, if they are non-zero, have positive real-parts iff and only iff there is a probability measure $\mu$ on the unit circle such that 
\begin{equation*}
\begin{split}
\; \mathop{{\rm \mathbb{M}in}}\limits_{\mu (\cdot)} {\rm \; }  \mathcal{D} \Bigg( \mathscr{F}_0\big(  \rho e^{j\vartheta}   \big) ,\int_0^{2 \pi} \frac{1-\rho^2}{1-2\rho cos (\vartheta-\vartheta^{\prime})+\rho^2} d \mu (\vartheta^{\prime})\Bigg),
\end{split}
\end{equation*}
holds, while
\begin{equation*}
\begin{split}
\; \mathscr{F}_0  =: d \Big(\Phi_1 \big(\mathscr{X}^{(t)}\big)\Big):= \mathscr{f}\big(\mathscr{e}, \mathscr{p}, t \big) d \mathscr{B}_0,
\end{split}
\end{equation*}
since it is proven that $\mathscr{F}_0$ is \textit{harmonic}, that is
\begin{equation*}
\begin{split}
\; \nabla^2 \mathscr{F}_0 = 0,
\end{split}
\end{equation*}
according to the Newton's second law of motion, roughly speaking.

The minimisation defined above can be re-written as
\begin{equation*}
\begin{split}
\; \mathop{{\rm \mathbb{M}in}}\limits_{\mu (\cdot)} {\rm \; }  \mathcal{D} \Bigg( \mathscr{F}_0\big(  \rho e^{j\vartheta}   \big) ,\int_0^{2 \pi} \frac{1+z(\vartheta)e^{-i \vartheta^{\prime}}}{1-z(\vartheta)e^{-i \vartheta^{\prime}}} d \mu (\vartheta^{\prime})\Bigg),
\end{split}
\end{equation*}
for the case where $\mathscr{F}_0$ is \textit{holomorphic}.

\textsc{\textbf{Remark 3.}} \textit{The term $ \frac{1+z(\vartheta)e^{-i \vartheta^{\prime}}}{1-z(\vartheta)e^{-i \vartheta^{\prime}}} $ can be re-expressed as
\begin{equation*}
\begin{split}
\;  \frac{1+z(\vartheta)e^{-i \vartheta^{\prime}}}{1-z(\vartheta)e^{-i \vartheta^{\prime}}}=e^{i \frac{\vartheta-\vartheta^{\prime}}{2}} \frac{e^{i \frac{\vartheta^{\prime}}{2}}+e^{-i \frac{\vartheta^{\prime}}{2}}}{e^{i \frac{\vartheta^{\prime}}{2}}-e^{-i \frac{\vartheta^{\prime}}{2}}} ,
\end{split}
\end{equation*}
while $z(\vartheta)=e^{i \vartheta}$. In this term, the term$\frac{e^{i \frac{\vartheta^{\prime}}{2}}+e^{-i \frac{\vartheta^{\prime}}{2}}}{e^{i \frac{\vartheta^{\prime}}{2}}-e^{-i \frac{\vartheta^{\prime}}{2}}} $ shows the story in relation to the tangent space expressed in Step 6 in Appendix \ref{sec:B}.}

The proof is now completed.$\; \; \; \blacksquare$

\section{Proof of Proposition 4}
\label{sec:D}
The achievable secret-key rate per dimension $\mathscr{R}_{key}$ is\footnote{see e.g. \cite{f5, f6}.}
\begin{equation*}
\begin{split}
\; \mathscr{R}_{key}=\mathop{{\rm \mathbb{E}}}\limits_{t} {\rm \; }  \Bigg \{  \big(1-\psi^{(t)} \big) \sigma^{(t)} \mathscr{I}^{(t)}_1 \Big( \frac{\omega(t)}{\sigma^{(t)} } \Big) + \cdots  \;\;\;\;\;\;\;\;\;\;\;\;\;\;\;\; \; \; \; \; \; \; \\ 
 \psi^{(t)}  \sigma^{(t)}    \bigg(    \mathscr{I}^{(t)}_1 \Big( \frac{\omega(t)}{\sigma^{(t)} }\Big)-\mathscr{I}^{(t)}_2 \Big( \frac{\omega(t)}{\sigma^{(t)} }      \Big)   \bigg) \Bigg \},
\end{split}
\end{equation*}
while
\begin{equation*}
\begin{split}
\; |\mathcal{S}^{(t)}_{ab}|=|\mathcal{S}^{(t)}_{ae}|=\sigma^{(t)},
\end{split}
\end{equation*}
holds, and $\omega(t)$ is the instantaneous signal-to-noise (SNR), as well as the fact that 
\begin{equation*}
\begin{split}
\;  \mathscr{I}^{(t)}_1 \Big(\omega(t)\Big) =: \mathcal{I} \big(  \mathscr{X}^{(t)};\mathscr{Y}^{(t)} | \mathcal{S}^{(t)}_{ab}   \big),
\end{split}
\end{equation*}
and
\begin{equation*}
\begin{split}
\;  \mathscr{I}^{(t)}_2 \Big(\omega(t)  \Big) =: \mathcal{I} \big(  \mathscr{X}^{(t)};\mathscr{Z}^{(t)} | \mathcal{S}^{(t)}_{ab}   \big),
\end{split}
\end{equation*}
hold\footnote{see e.g. \cite{f5, f6}.}, while $\mathscr{Z}$ is the observed data by Eve.

Now, one can assert that the Problem $\mathscr{P}_2$ expressed in Eqn. (\ref{eq:6}) can be re-written as the Problem $\mathscr{P}_3$ expressed in Eqn. (\ref{eq:7}) or the Problem $\mathscr{P}^{\prime}_{3}$ expressed in Eqn. (\ref{eq:8}), while for the $t-$th instant and $j-$th sub-channel, we have
\begin{equation*}
\begin{split}
\; \mathcal{P}r \big( j \in \mathcal{S}^{(t)}_{ae}| j \in \mathcal{S}^{(t)}_{ab}\big)=\psi^{(t)}, \forall j \in \mathbb{N},
\end{split}
\end{equation*}
and we principally define the following probabilistic constraint
\begin{equation*}
\begin{split}
\;  \mathcal{P}r \Big (\psi^{(t)} \big(  \mathscr{X}^{(t)}_0   \big) \ge \lambda^{(t)}_4 \Big) \le \lambda^{(t)}_5,
\end{split}
\end{equation*}
which can be in parallel with
 \begin{equation*}
\begin{split}
\;  \mathcal{P}r \Big (\omega^{(t)} \big(  \mathscr{X}^{(t)}_0   \big) \ge \lambda^{(t)}_6 \Big) \ge \lambda^{(t)}_7,
\end{split}
\end{equation*}
w.r.t. the arbitrary thresholds $\lambda^{(t)}_4$, $\lambda^{(t)}_5$, $\lambda^{(t)}_6$ and $\lambda^{(t)}_7$, roughly speaking.

The two probabilistic constraints theoretically introduced here stand respectively for the following facts: (\textit{i}) the upper-bound of the probability value $\psi^{(t)}$ should be constrained through a threshold; (\textit{ii}) and the lower-bound of the instantaneous SNR $\omega^{(t)}$ should also be constrained through a threshold as well.

Instead of the two constraints given above, one can constrain the secret-key rate as

\begin{equation*}
\begin{split}
\;  \mathcal{P}r \Big (\mathscr{R}^{(t)}_{key} \big(  \mathscr{X}^{(t)}_0   \big) \ge \lambda^{(t)}_8 \Big) \ge \lambda^{(t)}_9,
\end{split}
\end{equation*}
w.r.t. the arbitrary thresholds $\lambda^{(t)}_8$ and $\lambda^{(t)}_9$.

\textsc{\textbf{Remark 4.}} \textit{It is absolutely interesting to note that although perturbations are considered in the literature mainly due to attacks, we do not ignore the term $\xi + \Delta \xi$ in Eqns. (\ref{eq:7}-\ref{eq:8}-\ref{eq:9}) in order to preserve the optimality and generality of the problem. }

\textsc{\textbf{Remark 5.}} \textit{The three probabilistic constraints newly declared above are relaxable according to the principle of the Concentration-of-Measure $-$ something that theoretically proves an exponential bound for the two aforementioned constraints\footnote{See e.g. \cite{c1, c2, c3} for further discussion.}. }

The proof is now completed.$\; \; \; \blacksquare$

\begin{IEEEbiography}[{\includegraphics[width=1in,height=1.25in,clip,keepaspectratio]{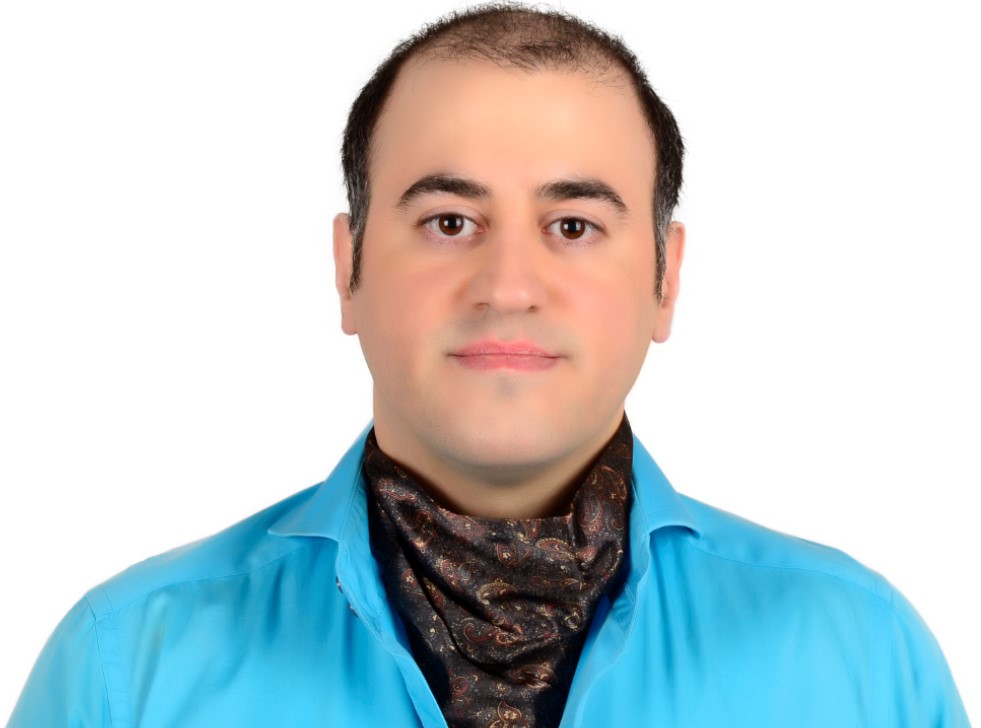}}]{Makan Zamanipour}
MAKAN ZAMANIPOUR (Researcher-ID: P-6298-2019; ORCID: 0000-0003-1606-9347; Scopus-ID: 56719734800) IEEE Member since 2015, born in Iran on 1983. His main research-field is Wireless communication theory, Information theory, Game theory and Optimisation. He has published a lot of papers in ISI-indexed journals as wll as reviewing for high-prestige ISI-indexed journals in IEEEs, Elsevier etc. His Google-Scholar profile and Publons are available online.
\end{IEEEbiography}
\markboth{IEEE, VOL. XX, NO. XX, X 2021}%
{Shell \MakeLowercase{\textit{et al.}}: Bare Demo of IEEEtran.cls for Computer Society Journals}
\end{document}